\documentclass[journal]{IEEEtran}
\usepackage{algorithm}
\usepackage{algorithmic}
\usepackage{bm}
\usepackage{url}
\usepackage{CJK}
\usepackage{multirow}
 \usepackage{amsfonts,amsmath,amssymb}
\usepackage{cite}
\usepackage{graphicx}
\usepackage{epstopdf}
\usepackage{xcolor}
\usepackage{caption}

\begin{document}
 \captionsetup[figure]{name={Fig.},labelsep=period}
\captionsetup{font={scriptsize}}

\title{User Association and Path Planning for UAV-Aided Mobile Edge Computing with Energy Restriction}
\author{Yuwen Qian, Feifei Wang, Jun Li, Long Shi, Kui Cai, and Feng Shu

\thanks{The work of J. Li and F. Shu was supported in part by National Key R$\&$D Program under Grants 2018YFB1004800, in part by National Natural Science Foundation of China under Grants 61727802, 61872184, and 6177124. The work of L. Shi and K. Cai was supported by SUTD-ZJU grant ZJURP1500102 and RIE2020 Advanced Manufacturing and Engineering (AME) programmatic grant A18A6b0057.

Y. Qian, F. Wang, J. Li and F. Shu are with School of Electronic and Optical Engineering, Nanjing University of Science and Technology, Nanjing, China (e-mail:\{admon, feifei.wang, jun.li, shufeng\}@njust.edu.cn). J. Li is also with the School of Computer Science and Robotics, National Research Tomsk Polytechnic University, Tomsk, 634050, Russia.

L. Shi and K. Cai are with Science and Math Cluster, Singapore University of Technology and Design, Singapore (e-mail:slong1007@gmail.com, cai\_kui@sutd.edu.sg).
} }

\maketitle

\begin{abstract}
Mobile edge computing (MEC) provides computational services at the edge of networks by offloading tasks from user equipments (UEs). This letter employs an unmanned aerial vehicle (UAV) as the edge computing server to execute offloaded tasks from the ground UEs. We jointly optimize user association, UAV trajectory, and uploading power of each UE to maximize sum bits offloaded from all UEs to the UAV, subject to energy constraint of the UAV and quality of service (QoS) of each UE. To address the non-convex optimization problem, we first decompose it into three subproblems that are solved with integer programming and successive convex optimization methods respectively. Then, we tackle the overall problem by the multi-variable iterative optimization algorithm. Simulations show that the proposed algorithm can achieve a better performance than other baseline schemes.
\end{abstract}

\begin{IEEEkeywords}
Mobile edge computing, UAV, user association, uploading power, trajectory optimization.
\end{IEEEkeywords}

\section{Introduction}\label{sec:1}
Mobile edge computing (MEC) has been emerging as a promising technology that enables computationally complex applications at resource-limited user equipments (UEs)~\cite{1}. By offloading resource-consuming tasks and caching popular resources at the edge servers, MEC can greatly alleviate computational burden on the UEs and reduce data processing delay~\cite{caching}. Furthermore, in rural and remote areas, it is not viable to deploy a large number of static servers for offloading tasks, due to complicated terrains and high costs. In these scenarios, mobile servers are more capable of combating uncertain environments than the static ones.

Among existing MEC strategies, the unmanned aerial vehicle (UAV) mounted MEC has attracted much attention because of its excellent maneuverability and low cost.
In~\cite{2}, the trajectory of a single UAV base station and user scheduling were jointly optimized to maximize the minimum average uploading rate of UEs, but it did not consider the battery capacities of UAVs. In practice, energy consumption of an UAV is a major issue when dealing with flight trajectory and offloaded tasks.
In light of this, a recent work~\cite{5} proposed an UAV-based MEC system that enables computing capabilities of the UAV to offer UEs with computation offloading services. This work aims at minimizing the overall energy consumption of the ground UEs under the UAV energy constraint by jointly optimizing task allocation and UAV trajectory.

In this letter, we study a novel UAV-aided MEC strategy to offer the ground UEs with efficient computational services.
Our goal is to maximize sum bits offloaded from all UEs to the UAV subject to UAV energy constraint and quality of service (QoS) requirement of each UE. Toward this end, we jointly optimize user association, UAV trajectory, and uploading power of each UE.
To solve this optimization problem, we decompose it into three separate subproblems, solve each subproblem by integer programming and successive convex optimization methods, and optimize the overall problem by applying multi-variable fixed iterative algorithm.
Simulation results not only show that the UAV with optimized trajectory can improve the sum bits, but also reveal the impact of design parameters on the system performance, thereby providing guideline for the UAV deployment in MEC.

\section{System Model and Problem Formulation}\label{sec:2}

\subsection{System Model}
\begin{figure}[t]
\centering
\includegraphics[width=2.5in]{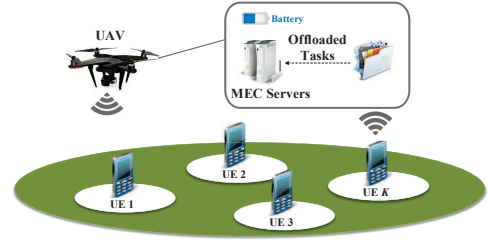}
\caption{The UAV-aided MEC system.}\label{Fig.1}
\end{figure}
As shown in Fig.~\ref{Fig.1}, we consider that a single UAV provides computational services for $K$ ground UEs, where the UAV flies over the UEs and receives the tasks. Meanwhile, the UAV with MEC servers executes these offloaded tasks from each UE and communicates with UEs to deliver the computational results. Also, UEs are distributed randomly each with a location $\mathbf {z}_k = (x_k,y_k), k \in \{1, 2,\cdots K\}$ over a $2$-D coordinate plane.

Let $T$ denote a task-offload period. During this period, the UAV departs from the origin, flies over the UEs for offloading tasks, and finally stops at a preset point $\mathbf{q}_T=(x_T,y_T)$. We divide $T$ into $N$ equal time slots with the duration of each time slot being $\delta_t=\frac{T}{N}$. Considering that each time slot is sufficiently small, the UAV location in the $n$-th time slot is ${\mathbf q}[n] = (x[n],y[n])$. Let {$\{{\mathbf q}\} = \{{\mathbf q}[n], \forall n\}$ denote a UAV trajectory over $N$ time slots with $\mathbf{q}[N] = \mathbf{q}_T$. In this letter, we only consider UAV flight altitude is fixed as a constant $H$. In addition, the UAV velocity $\nu[n]$ in the $n$-th time slot is constrained by the maximum velocity $\nu_{\max}$ as follows:
\begin{equation}\label{velocity_constraint}
\nu{[n]} = \frac{{{\Vert \mathbf q[n+1] - {\mathbf q}[n]\Vert}}}{{{\delta _t}}} \le {\nu _{\max }},\forall n.
\end{equation}

Let $b_k[n]$ be a binary variable of user association that indicates whether UE $k$ is served by the UAV in the $n$-th time slot. If $b_k[n] = 1$, the UAV receives workload from UE $k$ in the $n$-th time slot, otherwise $b_k[n] = 0$. At each time slot, the UAV serves at most one ground UE in the time division multiple access manner. These yield the following constraints:
\begin{equation}\label{user association constraint}
\sum\nolimits_{k = 1}^K {{b_k}[n]{\rm{ = 1}}} ,
{b_k}[n] \in \{ 0,1\},
\forall n, \forall k.\
\end{equation}

In this letter, we focus on the data uploading from the UEs to the UAV only and ignore latency caused by data downloading from the UAV to the UEs, as  the size of computational results from the UAV is much less than that of offloaded tasks from the UEs.
Consider that the channel between UAV and the UEs is line-of-sight (LoS)~\cite{2}. As such, the channel gain between UE $k$ and the UAV in the $n$-th time slot is given by
\begin{equation}\label{channel gain}
{h_k}[n] = \frac{{{\rho _0}}}{{{H^2} +  {\Vert \mathbf q[n] - {\mathbf z}_k\Vert}{^2}}},\forall n,
\end{equation}
where $\rho _0$ denotes the channel gain at the reference distance of 1m and the $3$-D distance from UE $k$ to the UAV is $\sqrt{{{{H^2} + {\rm{||{\mathbf q}[\emph{n}]}} - {{\rm{{\mathbf z}}}_k}{\rm{|}}{{\rm{|}}^2}}}}$. From~(\ref{channel gain}}), the uploading rate (\text {bits/s/Hz}) from UE $k$ to the UAV is given by
\begin{equation}\label{uploading rate}
{R_k}[n] = {\log _2}\left(1 + \frac{{{p_k}[n]{h_k}[n]}}{{{\sigma ^2}}}\right),\forall n, \forall k,
\end{equation}
where $\sigma ^2$ is the power of additive white Gaussian noise (AWGN) at the UAV and $p_k[n]$ is the uploading power of UE $k$ in the $n$-th time slot. Furthermore, we have
\begin{equation}\label{power sum}
{{\sum\nolimits_{n = 1}^N {{p_k}[n]} }{\delta_t}  \le {E_{\rm{U}}}, \forall k,}
\end{equation}
\begin{equation}\label{power minimum}
{p_k}[n] \ge {P_{\min }},\forall n, \forall k,
\end{equation}
where (\ref{power sum}) indicates that {{the total uploading energy of each UE over $N$ time slots}} is upper bounded by the constant ${E_{\rm{U}}}$ and (\ref{power minimum}) shows that the minimum uploading power ${P_{\min }}$ of each UE to support the transmission of basic information to the UAV. From~(\ref{uploading rate}), the bits offloaded from each UE to the UAV over $N$ time slots are given by 
\begin{equation}\label{Sk}
 {S_k} = \left(\sum\nolimits_{n = 1}^N {{b_k}[n]{R_k}[n]}\right){\delta _tB},\forall k,
\end{equation}
where $B$ is the channel bandwidth. To meet the quality of service (QoS) of each UE, $S_k$ yields
\begin{equation}\label{Dk}
{S_k} \ge {D_k},\forall k,
\end{equation}
where $D_k$ denotes the minimum offloaded bits of UE $k$ to guarantee its QoS. The energy consumption of the UAV is mainly caused by flight and computation. On one hand, with reference to~\cite{5}, the energy consumed by the UAV flight in the $n$-th time slot is ${e_{\rm{F}}}[n] = \kappa{{\nu^2[n]}}, \forall n$, where $\kappa = 0.5M\delta _t$ with $M$ being  the UAV weight. On the other hand, the computation energy for UE $k$ is given by
\begin{equation}\label{computation energy}
{e_k} = {\gamma _{\rm{C}}}{C_k}{S_k}{({f_{\rm{C}}})^2}, \forall k,
\end{equation}
where $C_k$ denotes the number of CPU cycles for computing every bit of UE $k$, $f_{\rm{C}}$ is the CPU frequency of the MEC servers, and $\gamma _{\rm{C}}$ is the effective switched capacitance~\cite{5}.

Consider that the energy limit of UAV battery is $E_0$. We restrict the total energy consumption comprising ${E_{\rm{F}}} = \sum\nolimits_{n = 1}^N {{e_{\rm{F}}}[n]}$ and ${E_{\rm{C}}} = \sum\nolimits_{k = 1}^K {{e_k}}$ to be ${E_{\rm{F}}} + {E_{\rm{C}}} \le {E_0}$.
\subsection{Problem Formulation}
Define $\hat{S}  = \sum\nolimits_{k = 1}^K {{S_k}}$ as the sum bits from all UEs to the UAV.
On one hand, achieving higher uploading rate motivates the UAV to fly closer to the UE, which in turn consumes more flight energy.
On the other hand, the more tasks the UEs offload, the more computation energy the UAV consumes.
Driven by these observations, our objective is to maximize the sum bits subject to battery energy and QoS constraints.
Towards this end, we jointly optimize the user association $\{b\}$, UAV trajectory $\{\mathbf q\}$, and UE uploading power $\{p\}$ as
\begin{subequations}\label{overall problem}
\begin{flalign}
\quad\mathcal P:&\quad{{\mathop {\max }\limits_{{{\{ b\} ,\{ {{\mathbf q}}\}, \{ p\}} }}\quad{\hat{S}}}}\\
\label{constrain_9b}
&~~\quad\quad{\text{s.t.}}~~~~\quad{S_k} \ge {D_k},\forall k,\\
\label{constrain_9c}
&~~~~~~\quad\quad\quad\quad{E_{\rm{C}}} + {E_{\rm{F}}} \le {E_0},\\
\label{constrain_9d}
&~~~~~~\quad\quad\quad\quad{\rm{{\mathbf q}[\emph{N}] = {\mathbf q}_\emph{T}}},\\
\label{constrain_9e}
&\quad\quad\quad\quad\quad\quad\eqref{velocity_constraint}, \eqref{user association constraint}, {\eqref{power sum}}, \text{and}~\eqref{power minimum},
\end{flalign}
\end{subequations}
where $\{b\} = \{b_k[n], \forall n, \forall k\}$ and $\{p\} = \{p_k[n], \forall n, \forall k\}$.

\section{Problem Optimization}\label{sec:3}
This section decouples {$\mathcal P$} into three subproblems and solves the problem by the multi-variable fixed iterative algorithm.
\subsection{User Association Optimization}\label{sec:3.1}
Given the optimized UAV trajectory and uploading power of each UE (see line 3 of Algorithm 1 in Section~\ref{sec:3.4}), this subproblem is to maximize the sum bits by optimizing the user association. As such, we can rewrite {$\mathcal P$} as
\begin{subequations}\label{user association optimizataion}
\begin{flalign}
\quad{\mathcal P_1}:\quad&\mathop{\max }\limits_{{{\{ {b}\} }}}\quad\hat{S} \\
\label{constrain_10a}
&~~\text{s.t.}\quad~\eqref{constrain_9b},~\eqref{constrain_9c},~\text {and}~\eqref{user association constraint}.
\end{flalign}
\end{subequations}
Note that {$\mathcal P_1$} is an integer programming problem. In this letter, we can solve $\mathcal P_1$ by branch and bound algorithm~\cite{wolseyinteger}.
\subsection{Trajectory Optimization}\label{sec:3.2}
This part focuses on the trajectory design based on the optimized user association and uploading power of each UE (see line 4 of Algorithm 1). First, we can rewrite {$\mathcal P$} as
\begin{subequations}\label{11}
\begin{flalign}
\label{constrain_11a}
\quad{\mathcal P_2}:\quad&\mathop {\max }\limits_{{{\{ {\mathbf q}\} }}}\quad\hat{S}\\
\label{constrain_11b}
&~~\text{s.t.}~\quad\eqref{constrain_9b},~\eqref{constrain_9c},~\eqref{constrain_9d},~\text {and}~\eqref{velocity_constraint}.
\end{flalign}
\end{subequations}
Note that {$\mathcal P_2$} is an intractable problem due to the non-convex expression ${R_k[n]}$ with respect to ${\mathbf q}[n]$. However, it is convex with respect to ${{\Vert \mathbf q[n] - {\mathbf z}_k\Vert}}^2$.
Let {{$\left\{{\mathbf q^{r}}\right\} = \{\mathbf q^r[n], \forall n\}$}} be the optimized trajectory in the $r$-th iteration (see Algorithm 1).
Then we can replace ${R_k[n]}$ with its first-order Taylor expansion at ${\mathbf{q}^r}[n]$:
\begin{align}\label{R low}
{R_k}[n] =&~{\log _2}\left(1 + \frac{\rho _0{{p_k[n]}} }{\sigma ^2({{H^2} + {\Vert \mathbf q[n] - {\mathbf z}_k\Vert}{^2} })}\right)\nonumber\\
\ge&~ A_k^r[n]\left({{\Vert \mathbf q[n] - {\mathbf z}_k\Vert}{^2}- {\Vert \mathbf {q}^r[n] - {\mathbf z}_k\Vert}{^2}}\right) + W_k^r[n]\nonumber\\
=&~R_k^{{\text {low}},r}[n],
\end{align}
where $W_k^r[n]$ and $A_k^r[n]$ are the coefficients of Taylor expansion of {$R_k[n]$} at ${\mathbf {q}^r}[n]$ with respect to ${{\Vert \mathbf q[n] - {\mathbf z}_k\Vert}}^2$ and the inequality holds since any convex function is globally lower-bounded by its first-order Taylor expansion at any point~\cite{2}.
Plugging~(\ref{R low}) into~(\ref{Sk}), we further have
\begin{equation}\label{13}
{S^{{\text {low}},r}_k} = {\sum\nolimits_{n = 1}^N \left({b_k}[n]R_k^{{\text {low}},r}[n]\right){\delta _t}B}
\end{equation}
as the lower-bound approximation of $S_k$ in~(\ref{constrain_9b}). Accordingly, $\hat{S}$ in~(\ref{constrain_11a}) can be replaced with $\hat{S}^{{\text {low}},r} = \sum\nolimits_{k = 1}^K {{S^{{\text {low}},r}_k}}$.

For the non-convex constraint of~(\ref{constrain_9c}), we derive a convex upper bound of non-convex function $R_k[n]$.
For any ${{\mathbf q^r}[n]}$, the following convex approximation function is a global upper bound of $R_k[n]$~\cite{7}, given by
\begin{align}\label{R up}
 R^{{\text {up}},r}_k[n] =&~W_k^r[n] + {(\nabla_{\mathbf q{[n]}}{R_k}[n])^{\mathcal T}_{\mathbf q^r[n]}}{({\mathbf q}[n] - {{\mathbf q}^r}[n])}\nonumber\\
&+\frac{{{L_{\nabla {R_k}[n]}}}}{2}{{\Vert \mathbf q[n] - {\mathbf q^r}[n]\Vert}^2},
\end{align}
where ${(\nabla_{\mathbf q{[n]}}{R_k}[n])_{\mathbf q^r[n]}}$ denotes the gradient of the non-convex function ${R_k}[n]$ at ${\mathbf {q}^r}[n]$ and $\mathcal {T}$ denotes the transpose.
Notably, ${(\nabla_{\mathbf q{[n]}}{R_k}[n])_{\mathbf q^r[n]}}$ is Lipschitz continuous with constant ${L_{\nabla {R_k}[n]}} = \max \{{\Vert{\nabla ^2}{R_k}[n]\Vert}_2\}$~\cite{bertsekas2015parallel}, where ${{\nabla ^2}}$ is Hessian matrix of ${R_k}[n]$ and ${\Vert{\cdot}\Vert}_2$ is the spectral norm of a matrix.
Plugging~(\ref{R up}) into~(\ref{Sk}) and~(\ref{computation energy}),  we can rewrite {$\mathcal P_2$} as
\begin{subequations}\label{22}
\begin{flalign}
\label{constrain_22p}
\quad{{\mathcal P_2^{'}}}:\quad&\mathop {\max }\limits_{{\{ {\mathbf q}\} }}\quad\hat{S}^{{\text {low}},r}\\
\label{constrain_22a}
&~~\text{s.t.}\quad~{S^{{\text {low}},r}_k} \ge D_k,\\
\label{constrain_22b}
&\quad\quad\quad{E_{\rm{C}}^{{\text {up}},r}} + {E_{\rm{F}}} \le {E_0},\\
\label{constrain_22c}
&\quad\quad\quad\eqref{constrain_9d},~\text {and}~\eqref{velocity_constraint},
\end{flalign}
\end{subequations}
where $E_{\rm{C}}^{{\text {up}},r}$ is the upper bound of $E_{\rm{C}}$ with respect to $R_k^{{\text {up}},r}[n]$. Therefore, {{{$\mathcal P_2^{'}$}}} becomes a convex problem, since~(\ref{constrain_22p}) and the left-hand-side (RHS) of~(\ref{constrain_22a}) are concave and the RHS of~(\ref{constrain_22b}) is convex.
Finally, we solve {{{$\mathcal P_2^{'}$}}} by standard convex optimization techniques~\cite{9}.

\subsection{Uploading Power Optimization}\label{sec:3.3}

Given the optimized user association and UAV trajectory (see line 5 of Algorithm 1), we can rewrite {$\mathcal P$} as
\begin{subequations}\label{power}
\begin{flalign}
\label{constrain powera}
\quad{\mathcal P_3}:\quad&\mathop {\max }\limits_{{{\{ {p}\} }}}\quad\hat{S}\\
\label{constrain powerb}
&~~\text{s.t.}~\quad\eqref{constrain_9b},~\eqref{constrain_9c},~{{\eqref{power sum}}}, \text{and}~\eqref{power minimum}.
\end{flalign}
\end{subequations}

Let $p_k^{r}[n]$ be the optimized uploading power in the $r$-th iteration. Similar to (13) in {$\mathcal P_2$}, we can replace $R_k[n]$ with its first-order Taylor expansion at $p_k^{r}[n]$, because any concave function is globally upper-bounded by its first-order Taylor expansion at any point~\cite{2}. Then, $E_{\rm{C}}$ can be replaced by its upper bound  ${E_{{\rm{C}},\text {power}}^{{\text {up}}, r}}$, which is a convex function with respect to $p_k^{r}[n]$. As such, we can rewrite {$\mathcal P_3$} as
\vspace{-0.1in}
\begin{subequations}\label{optimization of uploading power approximation}
\begin{flalign}
\label{constrain_uploading power approximation1}
\quad{\mathcal P_3^{'}}:\quad&\mathop {\max }\limits_{\{ {p}\} }\quad\hat{S}\\
\label{constrain_uploading power approximation2}
&~~\text{s.t.}~\quad{E_{{\rm{C}},\text {power}}^{{\text {up}}, r}} + {E_{\rm{F}}} \le {E_0},\\
\label{constrain_uploading power approximation13}
&~\quad\quad\quad\eqref{constrain_9b},{~\eqref{power sum}}, \text{and}~\eqref{power minimum},
\end{flalign}
\end{subequations}
 Note that {$\mathcal P_3^{'}$} is a convex problem and  we solve it by the standard convex optimization method~\cite{9}.

\subsection{Overall Optimization}\label{sec:3.4}
In this part, we solve the overall problem {$\mathcal P$} by the multi-variable fixed iterative algorithm as shown in Algorithm 1.
\vspace{-0.2in}
\begin{algorithm}
\caption{Multi-variable fixed iterative algorithm}
\begin{algorithmic}[1]
\STATE {Set $r = 0$ and the tolerance error $\epsilon$. Initialize UAV trajectory $\left\{{\mathbf q^0}\right\}$ and UE uploading power $\{p^0\}$}.
\REPEAT
    \STATE {Given $\left\{{\mathbf q^r}\right\}$ and $\{p^r\}$, find $\left\{{b^{r+1}}\right\}$ by solving {$\mathcal P_1$};}
    \STATE {Given $\left\{{b^{r+1}}\right\}$ and $\{p^r\}$, find $\left\{{\mathbf q^{r+1}}\right\}$ by solving {$\mathcal P_2^{'}$};}
    \STATE {Given $\left\{{b^{r+1}}\right\}$ and $\left\{{\mathbf q^{r+1}}\right\}$, find $\{p^{r+1}\}$ by solving {$\mathcal P_3^{'}$};}
    \STATE {Compute $\hat{S}^{r+1}$ using $\left\{{b^{r+1}}\right\}$, $\left\{{\mathbf q^{r+1}}\right\}$ and $\{p^{r+1}\}$.}
    \STATE {$r$ $\leftarrow$ $r + 1$.}
\UNTIL {$ |\hat{S}^{r+1} - \hat{S}^{r}  |\le \epsilon$}.
\end{algorithmic}
\end{algorithm}

\vspace{-0.1in}
\section{Numerical Results}\label{sec:4}

\begin{figure}[t]\vspace{-0.1in}\centering
\includegraphics[width=2.7in]{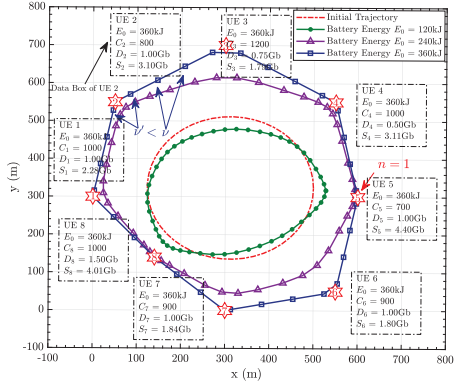}
\caption{Optimized trajectory of UAV-aided MEC with $f_{\rm{C}} = 2$GHz and $\nu_{\max} = 30$m/s.}
\label{Fig.2}\vspace{-0.15in}
\end{figure}

In this section, we evaluate performance of the proposed algorithm. We set a period of $T = 120\text s$ with $N = 50$ time slots. The number of the UEs is $K = 8$. The UAV altitude is $H = 50\text m$ and its weight is $M = 10\text k\text g$. The noise power is ${\sigma ^{\rm{2}}} = -140\text d\text B$. The channel bandwidth is $B = 10$MHz. The channel gain at the reference distance is  $\rho_0 = -50\text d\text B$.  The number of CPU cycles $C_k$ for each UE is shown in the data box of Fig.~\ref{Fig.2}.  {{We set {{$E_{\rm{U}} = $ 36J}} in (\ref{power sum}) and initial $p_k[n] =$ 0.3W in (\ref{power minimum}).}} The effective switched capacitance is $\gamma _{\rm{C}} = 10^{-27}$~\cite{10} and the initial trajectory is set as a ring.

Fig.~\ref{Fig.2} depicts the optimized trajectories under different battery energy, where the stars illustrate the locations of the UEs.
The data box shows the minimum offloaded bits $D_k$ and the bits $S_k$ from UE $k$ when battery energy $E_0 = 360$kJ.
First,   larger battery energy contributes to  larger UAV coverage area.
When $E_0$ is  small, e.g., $E_0 = 120$kJ, the UAV flies towards the UEs closer to the initial trajectory such as UEs 5 and 8, in order to save flight energy.
Conversely, when $E_0$ is sufficiently large, e.g., $E_0=360$kJ, the UAV can reach the location of each UE to obtain higher uploading rate.
Second, from the density of sampling points, we find that the UAV reduces the speed to collect more tasks as it flies closer to the UE. For example, the velocity $\nu'$ $<$ $\nu$, as it is observed that the UAV flies different distances over the same period of time.
Third, from $D_k$ and $S_k$ in the data box,  the UEs closer to the initial trajectory offload more bits than those farther from the trajectory. For example, ${S_7}$ $>$ ${S_6}$ as UE 7 is closer to the initial trajectory than UE 6. Fourth, we observe that  larger $C_k$ results in less offloaded bits due to  higher computational complexity. For example, $S_1<S_5$ as $C_1>C_5$.
\begin{figure}[t]\centering
\includegraphics[width=2.7in]{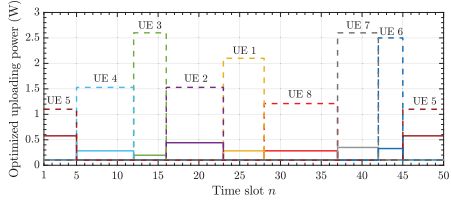}
\caption{Optimized uploading power $p_k[n]$ versus time slots, where the dashed and solid lines correspond to $E_0 = 360$kJ and $E_0 = 120$kJ respectively.}
\label{Fig.3}\vspace{-0.19in}
\end{figure}

Fig.~\ref{Fig.3} shows the optimized uploading power for all UEs in each time slot based on the optimized trajectory in Fig.~\ref{Fig.2}, where the dashed and solid lines correspond to $E_0 = 360$kJ and $E_0 = 120$kJ respectively. Lines with different colors represent different UEs.
First, the UE keeps the uploading power invariant when it is served by the UAV, while the UE maintains the minimum communication power $P_{\min} =$ 0.1W if it is not served. That is because the channel gain between the UAV and its served UE is almost unchanged during its service period.
Second, the uploading power of each UE increases as $E_0$ rises when it is served by the UAV, since each UE can upload more bits to the UAV with larger computation energy.
Third, UE $3$ has the highest uploading power when it is served among all UEs. This is due to the fact that UE 3 requires the largest computation energy for each bit (see $C_3$ in Fig.~\ref{Fig.2}) and its location is relatively farther from the initial trajectory.

\begin{figure}[t]
\centering
\includegraphics[width=2.7in]{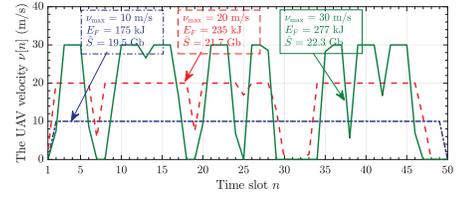}
\caption{The UAV velocity $\nu[n]$ versus time slots.}
\label{Fig.4}\vspace{-0.1in}
\end{figure}

Fig.~\ref{Fig.4} depicts the UAV velocity in each time slot versus maximum velocity under sufficient battery energy $E_0 = 360$kJ. The other parameters are the same with Fig.~\ref{Fig.2}.
First, when $\nu_{\max} = 10$m/s, the UAV needs to fly at this $\nu_{\max}$ over all time slots to serve each UE.
Second, when $\nu_{\max} = 30$m/s, the UAV first flies towards the UE quickly at this $\nu_{\max}$, and then it reduces speed to collect more tasks as it reaches the location of each UE as Fig.~\ref{Fig.2} shows. Third, from each data box, we observe that increasing maximum velocity not only increases sum bits $\hat{S}$ but also consumes more flight energy.
\begin{figure}[t]
\centering
\includegraphics[width=2.7in]{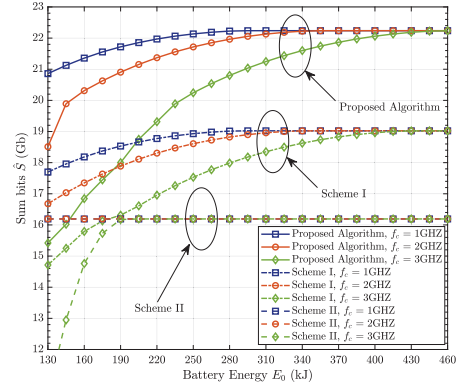}
\caption{The sum bits $\hat{S}$ versus battery energy $E_0$ with $D_k = 0.1\text {Gb}$.}
\label{Fig.5} \vspace{-0.21in}
\end{figure}

Fig.~\ref{Fig.5} examines the impact of battery energy $E_0$ on the sum bits $\hat{S}$, where Scheme I optimizes {$\mathcal P_1$} and {$\mathcal P_2$} under fixed UE uploading power~\cite{2}, and Scheme II  optimizes {$\mathcal P_1$} under fixed UAV trajectory and UE uploading power.
First, the proposed algorithm improves $\hat{S}$ by {16.8\%} and {37.3\%} compared with Scheme I and Scheme II respectively.
Second, $\hat{S}$ goes up as $E_0$ increases and $\hat{S}$ reaches the peak due to the limited uploading power. For example, for the proposed algorithm at $f_{\rm{C}} = 1$GHz, $\hat{S}$ rises as $E_0$ goes up and retains the peak at {$22.2$}Gb when $E_0\ge280$kJ.
Third, for any fixed battery energy, $\hat{S}$ drops as $f_{\rm{C}}$ increases. This is caused by the fact that the increase of $f_{\rm{C}}$ consumes more computing energy.

\section{Conclusion}\label{sec:5}
We have optimized user association, the trajectory of UAV and UE uploading power to maximize the sum bits of offloaded tasks in the UAV-aided MEC system subject to the UAV battery energy and QoS constraints, using integer programming and successive convex optimization methods.
Moving forward, it is of interest to study dynamic MEC systems where wireless channels, user locations, and UE tasks evolve in real time.

\bibliographystyle{IEEEtran}
\bibliography{Shi_WCL2019-0184_reference}

\begin{thebibliography}{1}
\providecommand{\url}[1]{#1}
\csname url@samestyle\endcsname
\providecommand{\newblock}{\relax}
\providecommand{\bibinfo}[2]{#2}
\providecommand{\BIBentrySTDinterwordspacing}{\spaceskip=0pt\relax}
\providecommand{\BIBentryALTinterwordstretchfactor}{4}
\providecommand{\BIBentryALTinterwordspacing}{\spaceskip=\fontdimen2\font plus
\BIBentryALTinterwordstretchfactor\fontdimen3\font minus
  \fontdimen4\font\relax}
\providecommand{\BIBforeignlanguage}[2]{{%
\expandafter\ifx\csname l@#1\endcsname\relax
\typeout{** WARNING: IEEEtran.bst: No hyphenation pattern has been}%
\typeout{** loaded for the language `#1'. Using the pattern for}%
\typeout{** the default language instead.}%
\else
\language=\csname l@#1\endcsname
\fi
#2}}
\providecommand{\BIBdecl}{\relax}
\BIBdecl

\bibitem{1}
Y.~Mao, C.~You, J.~Zhang, K.~Huang, and K.~B. Letaief, ``{A survey on mobile
  edge computing: The communication perspective},'' \emph{IEEE Commun. Surveys
  Tuts.}, vol.~19, no.~4, pp. 2322--2358, May 2017.

\bibitem{caching}
J.~{Li}, H.~{Chen}, Y.~{Chen}, Z.~{Lin}, B.~{Vucetic}, and L.~{Hanzo},
  ``Pricing and resource allocation via game theory for a small-cell video
  caching system,'' \emph{IEEE Journal on Selected Areas in Communications},
  vol.~34, no.~8, pp. 2115--2129, Aug 2016.

\bibitem{2}
Q.~Wu, Y.~Zeng, and R.~Zhang, ``{Joint trajectory and communication design for
  UAV-enabled multiple access},'' in \emph{Proc. IEEE Global Commun. Conf.},
  Singapore, Dec. 2017, pp. 1--6.

\bibitem{5}
S.~Jeong, O.~Simeone, and J.~Kang, ``{Mobile edge computing via a UAV-mounted
  cloudlet: Optimization of bit allocation and path planning},'' \emph{IEEE
  Trans. Veh. Technol.}, vol.~67, no.~3, pp. 2049--2063, Mar. 2018.

\bibitem{wolseyinteger}
L.~A. Wolsey, \emph{Integer Programming.}\hskip 1em plus 0.5em minus
  0.4em\relax Wiley, New York, 1998.

\bibitem{7}
G.~Scutari, F.~Facchinei, and L.~Lampariello, ``{Parallel and distributed
  methods for constrained nonconvex optimization part I: Theory},'' \emph{IEEE
  Trans. Signal Process.}, vol.~65, no.~8, pp. 1929--1944, Apr. 2017.

\bibitem{bertsekas2015parallel}
D.~Bertsekas and J.~Tsitsiklis, \emph{Parallel and Distributed Computation:
  Numerical Methods (Partial Solutions Manual)}.\hskip 1em plus 0.5em minus
  0.4em\relax Athena Scientific, 2015.

\bibitem{9}
S.~Boyd and L.~Vandenberghe, \emph{{Convex Optimization}}.\hskip 1em plus 0.5em
  minus 0.4em\relax Cambridge, U.K.: Cambridge Univ. Press, 2004.

\bibitem{10}
Y.~Mao, J.~Zhang, S.~H. Song, and K.~B. Letaief, ``{Stochastic joint radio and
  computational resource management for multi-user mobile edge computing
  systems},'' \emph{IEEE Trans. Wireless Commun.}, vol.~16, no.~9, pp.
  5994--6009, Jun. 2017.

\end{thebibliography}
\end{document}